\documentclass[hyper,letterpaper,12pt]{article}
\usepackage{a4wide}
\usepackage{amsmath}
\usepackage[
          colorlinks=true,
          linkcolor=blue,
          urlcolor=blue,
          filecolor=black,
          citecolor=blue,
]{hyperref}
\usepackage{color}
\usepackage{graphicx}
\usepackage{amsfonts}
\usepackage{amssymb}
\usepackage{epsfig}
\usepackage{subfigure}
\usepackage{latexsym}
\usepackage{mathrsfs}
\usepackage{cite}
\usepackage{multirow}
\usepackage{booktabs}
\usepackage{float}
\usepackage{multirow}
\begin{document}
	\title{Anisotropic quark stars in energy-momentum squared gravity with an interacting quark equation of state}
	\author{\large~~
		Qian Li$^1$$^{,}$\footnote{E-mail:qianli11090929@163.com}~,
        Xianming Liu$^{2}$$^{,}$$^{3}$$^{,}$\footnote{E-mail:xianmingliu@mail.bnu.edu.cn(corresponding author)}~
		Ruijun Tang$^1$$^{,}$\footnote{E-mail:RuijunTang875@163.com}~,
		Jiayong Xiao$^{1}$$^{,}$$^{4}$$^{,}$\footnote{E-mail:jyxiao@mails.ccnu.edu.cn}~
		\\
		\\
		\small $^1$ Department of Mathematics, Hubei Minzu University, Enshi Hubei 445000, China\\
        \small $^2$ Department of Physics, Hubei Minzu University, Enshi Hubei 445000, China\\
		\small $^3$  Key Laboratory of Quark and Lepton Physics(MOE), Institute of Particle Physics, \\
		\small Central China Normal University,Wuhan Hubei 430079, China\\
        \small $^4$  College of Physical Science and Technology, Central China Normal University, Wuhan 430079, China.\\}
		
	\maketitle
	
	\begin{abstract}
		\normalsize 
        We investigate anisotropic quark stars within the context of energy-momentum squared gravity (EMSG), assuming the internal composition consists of homogeneous, charge-neutral 3-flavor interacting quark matter with ${\cal O}(m_s^4)$ corrections. We performed numerical calculations of the equation of state (EoS) and the modified Tolman-Oppenheimer-Volkoff (TOV) structure equation. Additionally, we discussed the impact of the GB coupling parameter $\alpha $ on the mass-radius relationship. Furthermore, we compared our results with astronomical observational data. Our research offers a new perspective, proposing that in EMSG, accounting for anisotropy could lead to denser stars than in isotropic scenarios.
		
	\end{abstract}
	
	\newpage
	
	\section{Introduction}\label{sec:EMSG-g1}
	
	 General Relativity (GR) is a successful gravitational theory. However, it has certain limitations, such as its inability to explain the accelerated expansion of the universe and the singularity problem of the Big Bang, etc. To address these problems, researchers have proposed a series of modified gravity theories, such as $f(R)$ theory \cite{sotiriou2010f}, $f(R,T)$ theory \cite{harko2011f}, Einstein-Gauss-Bonnet (EGB) \cite{lanczos1938remarkable} theory, Rastall \cite{rastall1972generalization} theory, and others. Another covariant extension of GR is proposed in references \cite{katirci2014gravity,roshan2016energy}.  This theory introduces a nonlinear term ${T^2} = {T_{\mu \nu }}{T^{\mu \nu }}$ into the general action of GR, where ${T_{\mu \nu }}$ represents the energy-momentum tensor. Roshan and Shojai \cite{roshan2016energy} referred to this theory as energy-momentum squared gravity (EMSG). This matter-geometry coupling theory has been explored in cosmology and astrophysics, thus it is consistent with GR, and EMSG is equivalent to GR in vacuum. Roshan and Shojai applied EMSG to a homogeneous and isotropic spacetime, demonstrating the existence of a bounce during the early stages, thus resolving the singularity issue of the Big Bang. Board and Barrow \cite{board2017cosmological} conducted a study on EMSG, identifying numerous precise solutions for isotropic universes, and examining their characteristics in relation to early and late-time evolution, accelerated expansion, and the presence or prevention of singularities. The EMSG theory has been extensively applied in the study of cosmological solutions \cite{barbar2020viability}, charged black holes \cite{chen2020eikonal}, wormhole solutions \cite{moraes2018nonexotic, sharif2021viable}, ect.
  
	 A recent series of papers has sparked interest among researchers in the study of anisotropic stars. 
	 In the context of anisotropy, radial pressure and tangential pressure exhibit disparities. Ruderman \cite{ruderman1972pulsars} found theoretically that anisotropic effects may appear in stellar models when nuclear matter reaches densities higher than ${10^{15}}g/c{m^3}$. Herrera and Santos \cite{herrera1997local} speculated that anisotropic stars could exist within regions of strong gravity. Currently, many researchers have analyzed anisotropy pressure in the context of compact stars \cite{mak2003anisotropic, herrera2013newtonian, herrera2013general, herrera2014conformally, setiawan2019anisotropic, rizaldy2019neutron, maurya2019anisotropic, tangphati2021anisotropic, pretel2022anisotropic, tangphati2023criteria}. Additionally, the EoS plays a pivotal role in stellar investigations, as the mass-radius relationship of stars is inherently linked to the EoS. When exploring anisotropic stars, we will employ a more comprehensive the EoS, one proposed by Becerra-Vergara and colleagues \cite{becerra2019anisotropic}.
	 
	 In this paper, we employ numerical techniques to analyze the internal structure and physical properties of quark stars within the framework of EMSG. Additionally, we have compared our computational results with those of PSR J0348+0432 \cite{antoniadis2013massive}, PSR J1614-2230 \cite{demorest2010two}, PSR J0437-4715 \cite{verbiest2008precision}, PSR B1913+16 \cite{weisberg2005binary}. The structure of this paper is as follows: in section \ref{sec:EMSG-g2}, we give the field equations for the EMSG and derive the equations of motion. Additionally, We derive the TOV equations for stars under static and spherically symmetric metric conditions. In section \ref{sec:EMSG-g3}, we specifically define the EoS for interacting quark matter. In section \ref{sec:EMSG-g4}, we employed numerical methods to solve these equations and presented our significant findings. In section \ref{sec:EMSG-g5},we employed analytical methods and graphical representations to describe the physical properties of the quark star model. In section \ref{sec:EMSG-g6}, we provided a summary of our research results.

	\section{Field equations in energy-momentum squared gravity}
	\label{sec:EMSG-g2}
     In this section, we introduce fundamental aspects of energy-momentum squared gravity (EMSG) theory. We set $G = c = 1$. The action of EMSG \cite{katirci2014gravity, roshan2016energy, tangphati2022mass, singh2021color} is 
	\begin{equation}
	S = \int {(\frac{1}{{2k}}{\rm{{\cal R}}}}  + \alpha {T_{\mu \nu }}{T^{\mu \nu }} + {{\cal L}_m})\sqrt { - g} {d^4}x \label{Eq1}
	\end{equation}
	where ${\rm{{\cal R}}}$ is the Ricci scalar, $k = 8\pi$, $\alpha$ is the coupling constant, ${T_{\mu \nu }}$ is the energy-momentum tensor, ${{\cal L}_m}$ is the matter Lagrangian density.
	
	The energy-momentum tensor is often defined as
	\begin{equation}
	{T_{\mu \nu }} =  - \frac{2}{{\sqrt { - g} }}\frac{{\delta (\sqrt { - g} {{\cal L}_m})}}{{\delta {g^{\mu \nu }}}} \label{Eq2}
	\end{equation}
	
	If ${{\cal L}_m}$ is determined solely by the metric components and not by their derivatives, we have 

	\begin{equation}
	{T_{\mu \nu }} = {{\cal L}_m}{g_{\mu \nu }} - 2\frac{{\partial {{\cal L}_m}}}{{\partial {g^{\mu \nu }}}} \label{Eq3}
	\end{equation}
	 
	The following gravitational field equations is obtained by varying the action give in Eq.\eqref{Eq1} with regard to the metric
	
	\begin{equation}
    {{\rm{{\cal G}}}_{\mu \nu }} = 8\pi {T_{\mu \nu }} + 8\pi \alpha ({g_{\mu \nu }}{T_{\beta \gamma }}{T^{\beta \gamma }} - 2{\Theta _{\mu \nu }}) \label{Eq4}
	\end{equation}
	where ${{\rm{{\cal G}}}_{\mu \nu }} = {{\rm{{\cal R}}}_{\mu \nu }} - \frac{1}{2}{g_{\mu \nu }}{\rm{{\cal R}}}$ is the Einstein tensor. The tensor ${\Theta _{\mu \nu }}$ as
	
	\begin{equation}
	{\Theta _{\mu \nu }} = {T^{\beta \gamma }}\frac{{\delta {T_{\beta \gamma }}}}{{\delta {g^{\mu \nu }}}} + {T_{\beta \gamma }}\frac{{\delta {T^{\beta \gamma }}}}{{\delta {g^{\mu \nu }}}} =  - 2{{\cal L}_m}({T_{\mu \nu }} - \frac{1}{2}{g_{\mu \nu }}T) - T{T_{\mu \nu }} + 2T_\mu ^\gamma {T_{\nu \gamma }} - 4{T^{\beta \gamma }}\frac{{{\partial ^2}{{\cal L}_m}}}{{\partial {g^{\mu \nu }}\partial {g^{\beta \gamma }}}} \label{Eq5}
	\end{equation}
	where $T = {g^{\mu \nu }}{T_{\mu \nu }}$ is the trace of the energy-momentum tensor.
	
	The energy momentum tensor can be written as 
	
	\begin{equation}
	{T_{\mu \nu }} = (\rho  + {P_ \bot }){u_\mu }{u_\nu } + {P_ \bot }{g_{\mu \nu }} + ({P_r} - {P_ \bot }){\chi _\mu }{\chi _\nu } \label{Eq6}
	\end{equation}
	where $\rho$ is the energy density, $P_r$ is the radial pressure,${P_ \bot }$is the tangential pressure, ${u^\mu }$ is the four-velocity  of the fluid, ${\chi ^\mu }$ is the space-like unit vector in the radial direction. 	${u_\mu }{u^\mu } =  - 1$, ${\nabla _\nu }{u^\mu }{u_\mu } = 0$ and ${\chi _\mu }{\chi ^\mu } = 1$ such that  ${u^\mu } = \left( {\begin{array}{*{20}{c}}
		{\begin{array}{*{20}{c}}
			{\frac{1}{{\sqrt {{e^{2\nu}}} }}}&0
			\end{array}}&{\begin{array}{*{20}{c}}
			0&0
			\end{array}}
		\end{array}} \right)$, ${\chi^\mu } = \left({\begin{array}{*{20}{c}}
		{\begin{array}{*{20}{c}}
			{\frac{1}{{\sqrt {{e^{2\lambda }}} }}}&0
			\end{array}}&{\begin{array}{*{20}{c}}
			0&0
			\end{array}}
		\end{array}} \right)$. 
	
	As is known, the definition of the matter Lagrangian giving the perfect fluid energy-momentum tensor is not unique, and we assume that ${{\cal L}_m} = P = \frac{1}{3}({P_r} + 2{P_ \bot })$  \cite{singh2020physical, maurya2020anisotropic}. We note that the EMT given in Eq.\eqref{Eq3} does not include the second variation of ${{\cal L}_m}$, hence the last term of Eq.\eqref{Eq5} is null. The conservation equation can be found by covariant derivative of Eq.\eqref{Eq4}, which gives
	
	\begin{equation}
	{\nabla ^\mu }{T_{\mu \nu }} =  - \alpha {g_{\mu \nu }}{\nabla ^\mu }({T_{\beta \gamma }}{T^{\beta \gamma }}) + 2\alpha {\nabla ^\mu }{\Theta _{\mu \nu }} \label{Eq7}
	\end{equation}
	where we can see that the conservation of local covariant energy-momentum is rarely satisfied, i.e. ${\nabla ^\mu }{T_{\mu \nu }} \ne 0$. But in the case of $\alpha  = 0$ is conservative.
		
	We adjust the field equations of EMSG from Eq.\eqref{Eq4} to the GR field equation as follows \cite{tangphati2023criteria}
	
	\begin{equation}
	{{\rm{{\cal G}}}_{\mu \nu }} = 8\pi {T_{\mu \nu }}_{(eff)} \label{Eq8}
	\end{equation}
	where ${T_{\mu \nu }}_{(eff)}$ is the effective energy-momentum tensor. This leads to a new version of the conservation law of the energy-momentum tensor of EMSG as
	
	\begin{equation}
	{\nabla ^\mu }{T_{\mu \nu }}_{(eff)} = 0 \label{Eq9}
	\end{equation}
	
	The effective energy momentum tensor can be written as
		
	\begin{equation}
	{T_{\mu \nu }}_{(eff)} = ({\rho _{(eff)}} + {P_ \bot }_{(eff)}){u_\mu }{u_\nu } + {P_ \bot }_{(eff)}{g_{\mu \nu }} + ({P_r}_{(eff)} - {P_{ \bot (eff)}}){\chi _\mu }{\chi _\nu } \label{Eq10}
	\end{equation}
	where ${\rho _{(eff)}}$ is the effective energy density, ${P_r}_{(eff)}$ is the effective radial pressure, ${P_ \bot }_{(eff)}$ is the effective tangential pressure. 
	
	Now, substituting Eq.\eqref{Eq10} in Eq.\eqref{Eq8}, the field equation becomes
	
	\begin{equation}
	{{\rm{{\cal G}}}_{\mu \nu }} = 8\pi ({\rho _{(eff)}} + {P_ \bot }_{(eff)}){u_\mu }{u_\nu } + {P_ \bot }_{(eff)}{g_{\mu \nu }} + ({P_r}_{(eff)} - {P_{ \bot (eff)}}){\chi _\mu }{\chi _\nu } \label{Eq11}
	\end{equation}
	
	 where ${\rho _{(eff)}}$, ${P_r}_{(eff)}$, and ${P_ \bot }_{(eff)}$ can be written in terms of the traditional $\rho$, ${P_r}$, and ${P_ \bot }$ as follows

	\begin{equation}
	{\rho _{(eff)}} = \rho  + \alpha (3{P_r}^2 + 4{P_r}\sigma  + \frac{{2{\sigma ^2}}}{3} + (8{P_r} + \frac{{16\sigma }}{3})\rho  + {\rho ^2}) \label{Eq12}
	\end{equation}
	
	\begin{equation}
	{P_r}_{(eff)} = {P_r} + \alpha (3{P_r}^2 + \frac{8}{3}{P_r}\sigma - \frac{{2{\sigma ^2}}}{3} + \frac{4}{3}\sigma \rho  + {\rho ^2}) \label{Eq13}
	\end{equation}
	
	\begin{equation}
	{P_ \bot }_{(eff)} = {P_r} + \sigma  + \alpha (3{P_r}^2 + \frac{{14}}{3}{{\mathop{\rm P}\nolimits} _r}\sigma  + 2{\sigma ^2} - \frac{2}{3}\sigma \rho  + {\rho ^2}) \label{Eq14}
	\end{equation}
	where $\sigma  = {P_ \bot } - {P_r} $.

	We consider a spherically symmetric and static metric in the form
	
	\begin{equation}
	d{s^2} =  - {e^{2\nu }}d{t^2} + {e^{2\lambda }}d{r^2} + {r^2}(d{\theta ^2} + {\sin ^2}\theta d{\phi ^2}) \label{Eq15}
	\end{equation}
	there are two independent functions $\nu (r)$ and $\lambda (r)$ in Eq.\eqref{Eq15}. 

	Using Eq.\eqref{Eq11} and the metric given in Eq.\eqref{Eq15}, we obtain following the set of equations 
	
	\begin{equation}
    \frac{{{e^{ - 2\lambda }}}}{{{r^2}}}(2r\lambda ' - 1) + \frac{1}{{{r^2}}} = 8\pi {\rho _{(eff)}} \label{Eq16}
	\end{equation}
	\begin{equation}
	\frac{{{e^{ - 2\lambda }}}}{{{r^2}}}(2r\nu ' + 1) - \frac{1}{{{r^2}}} = 8\pi {P_{r(eff)}} \label{Eq17}
	\end{equation}
	\begin{equation}
	\frac{{{e^{ - 2\lambda }}}}{r}(r\nu {'^2} - \lambda  + \nu (1 - r\lambda ') + r\nu '') = 8\pi {P_{ \bot (eff)}} \label{Eq18}
	\end{equation}
	
	To solve Eqs.\eqref{Eq16}-\eqref{Eq18}, we define the relation between the parameters $\lambda$, mass $m(r)$, and radius $r$
	
	\begin{equation}
	{e^{ - 2\lambda }} = 1 - \frac{{2m(r)}}{r} \label{Eq19}
	\end{equation}
	
	The metric function $\nu(r)$ and the effective radial pressure ${P_{r(eff)}}$ relation formula is
	
	\begin{equation}
	\frac{{d\nu}}{{dr}} = \frac{{ - 2{P_{r(eff)}} + 2{P_{ \bot (eff)}} - r{P_{r(eff)}}'}}{{r({P_{r(eff)}} + {\rho _{(eff)}})}} \label{Eq20}
	\end{equation}

	Using Eqs.\eqref{Eq13}, \eqref{Eq17}, \eqref{Eq19}, and \eqref{Eq20}, we obtain the following equation
		
	\begin{equation}
	\frac{{d{P_r}_{(eff)}}}{{dr}} = \frac{{(2r - 4m){\sigma _{(eff)}}}}{{r(r - 2m)}} - \frac{{m{P_{r(eff)}} + 4\pi {r^3}{P_{r(eff)}}^2 + m{\rho _{(eff)}} + 4\pi {r^3}{P_{r(eff)}}{\rho _{(eff)}}}}{{r(r - 2m)}} \label{Eq21}
	\end{equation}

	Using Eqs.\eqref{Eq12}, \eqref{Eq13}, \eqref{Eq16}, \eqref{Eq19}, and \eqref{Eq21}, we find that the modified Tolman-Oppenheimer-Volkoff (TOV) equations, now read
	
	\begin{equation}
	m'(r) = 4\pi {r^2}(\rho  + \alpha (3{P_r}^2 + {\rho ^2} + 4{P_r}\sigma  + \frac{{2{\sigma ^2}}}{3} + \rho (8{P_r} + \frac{{16\sigma }}{3}))) \label{Eq22}
	\end{equation}
	and
	\begin{equation}
	\begin{split}
	{P_r}'(r) = & (\rho {( - 3r(r - 2m))^{ - 1}}(1 + \frac{{{P_r}}}{\rho })(3 + 2\alpha (3\rho  + 9\rho \frac{{{P_r}}}{\rho } + 10\sigma ))\\&(3m + 4\pi {r^3}(3\rho \frac{{{P_r}}}{\rho } + \alpha (3{\rho ^2}(1 + 3{{\frac{{{P_r}}}{\rho }}^2})) + 4\rho \sigma (1 + 2\frac{{{P_r}}}{\rho }) - 2{\sigma ^2}))\\&+ (6\sigma  + 2\alpha (6{P_r}\sigma  + 8{\sigma ^2} - 6\sigma \rho )){r^{ - 1}} - 4\alpha \sigma '(r)(\rho  + 2\frac{{{P_r}}}{\rho }\rho  - \sigma ))\\&{(3 + 2\alpha (9\rho \frac{{{P_r}}}{\rho } + 4\sigma  + 2\frac{{d\rho }}{{d{P_r}}}(3\rho  + 2\sigma )))^{ - 1}} \label{Eq23}
	\end{split}
	\end{equation}
	
	Based on observational measurements of neutron stars, it has been shown in \cite{tangphati2022mass, akarsu2018constraint} that the parameter $ \alpha$ has been constrained, $- 5.60763 \times {10^{ - 5}} \le \alpha  \le 2.08283 \times {10^{ - 4}}f{m^3}/MeV$. We will attempt to obey the range in studying quark stars in EMSG with quark matter EoS.
	\section{Quark matter equation of state (EoS) and the boundary conditions of the quark star}
	\label{sec:EMSG-g3}
	
   In order to address the system of Eqs.\eqref{Eq22} and \eqref{Eq23}, it is necessary to define an EoS and introduce an anisotropy function $\sigma$ due to the presence of an additional degree of freedom ${P_ \bot }$. For the investigation of quark stars within the framework of EMSG, we employ the interacting quark equation of state denoted as ${P_r}(\rho )$, which is described as follows: \cite{becerra2019anisotropic}
	
	\begin{equation}
	{P_r} = \frac{1}{3}(\rho  - 4B) - \frac{{m_s^2}}{{3\pi }}\sqrt {\frac{{\rho  - B}}{{{a_4}}}}  + \frac{{m_s^4}}{{12{\pi ^2}}}[1 - \frac{1}{{{a_4}}} + 3\ln (\frac{{8\pi }}{{3m_s^2}}\sqrt {\frac{{\rho  - B}}{{{a_4}}}} )] \label{Eq24}
	\end{equation}
	where $\rho$ is the energy density of homogeneously distributed quark matter (also to ${\rm{{\cal O}}}(m_s^4)$ in the Bag model), B is the bag constant, $57MeV/f{m^3} \le B \le 92MeV/f{m^3}$  \cite{fiorella2018nuclear, blaschke2018phases}, $m_s$ is the quark strange mass, the interacting parameter $a_4$ comes from QCD corrections on the pressure of the quark-free Fermi sea, and it is directly related with the mass-radius relations of quark stars. 
	
	For the anisotropic quark matter, the EoS employed in this work was proposed by Becerra-Vergara et al \cite{becerra2019anisotropic}. 
	
	\begin{equation}
	\begin{array}{l}
	\begin{aligned}
	{P_ \bot } = &{P_c} + \frac{1}{3}(\rho  - 4{B_ \bot }) - \frac{{m_s^2}}{{3\pi }}\sqrt {\frac{{\rho  - {B_ \bot }}}{{a_4^ \bot }}}  + \frac{{m_s^4}}{{12{\pi ^2}}}[1 - \frac{1}{{a_4^ \bot }} + 3\ln (\frac{{8\pi }}{{3m_s^2}}\sqrt {\frac{{\rho  - {B_ \bot }}}{{a_4^ \bot }}} )]\\& - \frac{1}{3}({\rho _c} - 4{B_ \bot }) + \frac{{m_s^2}}{{3\pi }}\sqrt {\frac{{{\rho _c} - {B_ \bot }}}{{a_4^ \bot }}}  - \frac{{m_s^4}}{{12{\pi ^2}}}[1 - \frac{1}{{a_4^ \bot }} + 3\ln (\frac{{8\pi }}{{3m_s^2}}\sqrt {\frac{{{\rho _c} - {B_ \bot }}}{{a_4^ \bot }}} )]
	\end{aligned}
	\end{array} \label{Eq25} 
	\end{equation}
	where $P_c$ is the central radial pressure, $\rho$ is the central energy density. Note that at the center of the star i.e., r = 0 the radial and tangential pressures are the same for $B = B_ \bot$ and $a_4 = a_4^ \bot$. This condition represents the case of an isotropic fluid and desirable for any anisotropic fluid sphere. It is worth mentioning that $B_ \bot$ and $a_4^ \bot$ parameters are the contributions on the tangential component of the pressure, and run in the same range of values as B and $a_4$.  
	
	In order to unravel the internal structure of quark stars in EMSG, we performed numerical calculations of the modified TOV equations Eqs.\eqref{Eq22} and \eqref{Eq23} and EoS Eqs.\eqref{Eq24} and \eqref{Eq25}. At the center of the quark star, we consider the following boundary conditions
	
	\begin{equation}
	\rho (0) = {\rho _c},  m(0) = 0,  {P_r}({R}) = {P_ \bot }({R}) = 0 \label{Eq26}      
	\end{equation}
	where ${\rho _c}$ is the central energy density, ${R}$ is the radius of the star.
	
	\section{Numerical details and analysis of the mass-radius relations}
	\label{sec:EMSG-g4}
	
	In this section, we present the detailed outcomes of the EoS Eqs.\eqref{Eq24} and \eqref{Eq25} and showcase all the relevant outcomes obtained from investigating anisotropic quark stars within the framework of 4D EMSG theory. To solve the field equations, we will directly employ the TOV equation Eq.\eqref{Eq23} along with the mass function equation Eq.\eqref{Eq22}. It is worth noting that these two equations involve four unknown quantities, namely $\rho (r)$, ${P_ \bot }(r)$, ${P_r}(r)$, and $m(r)$. Therefore, it is not possible to solve these two equations simultaneously. Subsequently, we considered the inclusion of anisotropic matter in the EoS Eqs.\eqref{Eq24} and \eqref{Eq25}. Since the system of equations Eqs.\eqref{Eq22}-\eqref{Eq25} lacks an analytical solution, we numerically solved this system using boundary conditions Eq.\eqref{Eq26}. We also examined the impact of the theory's free parameters on the characteristics of anisotropic quark stars. In this study, the free parameters under consideration are $\alpha $, $B$, ${a_4}$, ${B_ \bot }$, and $a_4^ \bot $. The mass of quark stars is expressed in units of solar mass ${M_ \odot }$, the radius of quark stars is measured in $km$, $\rho (r)$, ${P_ \bot }(r)$, ${P_r}(r)$, ${B_ \bot }$, and $B$ are in units of $MeV/f{m^3}$, ${m_s}$ is in units of $MeV$, and $\alpha $ is in units of $f{m^3}/MeV$. ${a_4}$ and ${a_{4 \bot }}$ are dimensionless.
	
	We have plotted the $M-R$ relationship graph, incorporating the observed masses of the following pulsars: PSR J0348+0432 has a mass of $2.01 \pm 0.04{M_ \odot }$ \cite{antoniadis2013massive}, PSR J1614-2230 has a mass of $1.97 \pm 0.04{M_ \odot }$ \cite{demorest2010two}, PSR J0437-4715 has a mass of $1.76 \pm 0.20{M_ \odot }$ \cite{verbiest2008precision}, and PSR B1913+16 has a mass of $1.4408 \pm 0.008{M_ \odot }$ \cite{weisberg2005binary}.
	
	\begin{figure}
		\centering
		\subfigure{
		\includegraphics[width=0.5\textwidth]{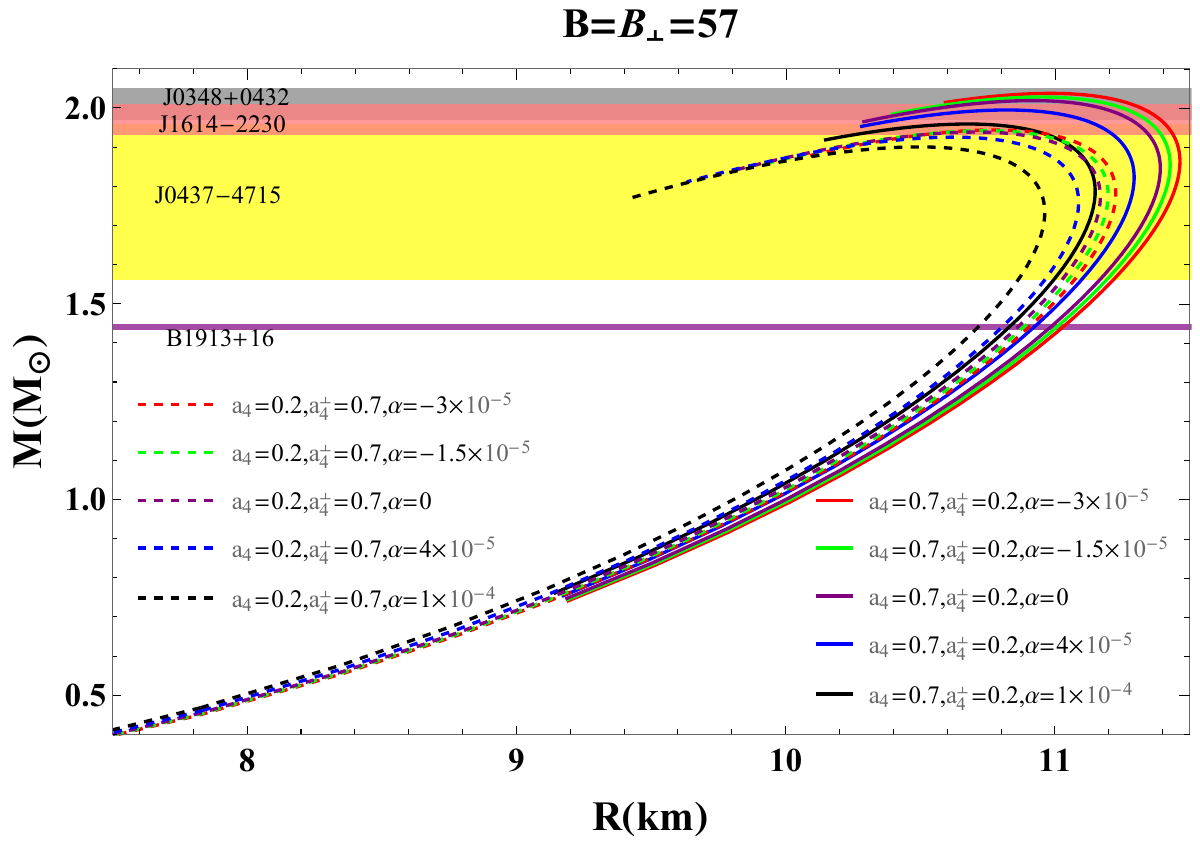}
		\includegraphics[width=0.5\textwidth]{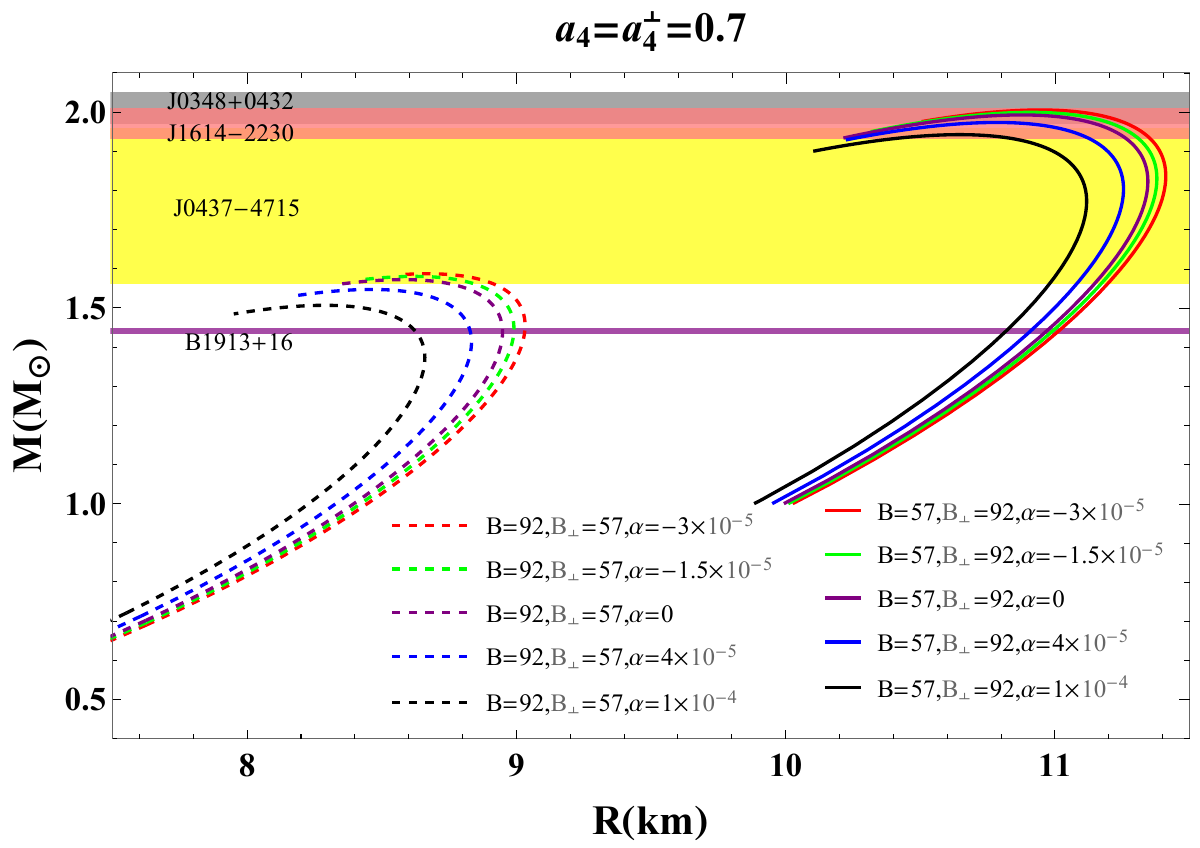}
		}
	    \caption{We plotted the mass-radius relationship for anisotropic quark stars. $B = {B_ \bot } = 57MeV/f{m^3}$, ${a_4} \ne a_4^ \bot $(left panel) and ${a_4} = a_4^ \bot  = 0.7$, $B \ne {B_ \bot }$ (right panel). The horizontal region displays measurement data from various pulsars: PSR J0348+0432 (gray) \cite{antoniadis2013massive}, PSR J1614-2230 (pink) \cite{demorest2010two}, PSR J0437-4715 (yellow) \cite{verbiest2008precision}, and PSR B1913+16 (purple) \cite{weisberg2005binary}.}\label{fig:1}
	\end{figure}
	
	In Figure \ref{fig:1} (left panel), we set the bag constant to $B = {B_ \bot } = 57MeV/f{m^3}$ and take values for $\alpha $, ${a_4}$, and $a_4^ \bot $ over a range. In the real curve part, we take ${a_4} = 0.7$, $a_4^ \bot  = 0.2$, and $\alpha$ takes several values. The maximum mass of the star decreases as $\alpha$ increases. In the imaginary curve part, we take ${a_4} = 0.2$, $a_4^ \bot  = 0.7$, and $\alpha$ takes several values. Similar to the real curve part, the maximum mass of the star decreases as $\alpha$ increases. From the left panel of Figure \ref{fig:1} we find that only the real curve falls within the mass limit $2.01 \pm 0.04{M_ \odot }$ of the pulsar J0348+0432. Our observations are also in good agreement with the detected pulsar J1614-2230.
	
	In Figure \ref{fig:1} (right panel), we set ${a_4} = a_4^ \bot  = 0.7$ and take values for $\alpha$, $B$ and ${B_ \bot }$ in a range. In the real curve part, we take $B = 57MeV/f{m^3}$, ${B_ \bot } = 92MeV/f{m^3}$, and $\alpha$ takes several values. Similar to the left panel, the maximum mass of the star decreases with increasing $\alpha$. In the imaginary curve part, we take $B = 92MeV/f{m^3}$, ${B_ \bot } = 57MeV/f{m^3}$, and $\alpha$ takes several values. From Figure \ref{fig:1} (right panel) we can see that only the real curve falls within the mass limit $1.76 \pm 0.20{M_ \odot }$ of pulsar J1614-2230. Our observations are also in good agreement with the detected pulsar J0437-4715.
	
	In Table \ref{tab:1}, we have set $\alpha  =  - 3 \times {10^{ - 5}}f{m^3}/MeV$, ${m_s} = 50MeV$, and varied the values of $B = {B_ \bot }$, ${a_4}$ and $a_4^ \bot $  within a specified range. This consequently yields data results for the maximum mass and its corresponding radius, central energy density, compactness, and gravitational redshift. In the EMSG, anisotropic stars can reach a maximum mass of 1.81-1.87${M_ \odot }$ when $\alpha  =  - 3 \times {10^{ - 5}}f{m^3}/MeV$, $B = {B_ \bot } = 70MeV/f{m^3}$, ${m_s} = 50MeV$, and ${a_4} = 0.7$, while isotropic stars at $\alpha  =  - 4.80654 \times {10^{ - 5}}f{m^3}/MeV$ have a maximum mass of 1.80${M_ \odot }$ \cite{tangphati2022mass}. Therefore, considering anisotropic factors, the mass of the star is greater.  It can be observed from Table 1 that when $\alpha  =  - 3 \times {10^{ - 5}}f{m^3}/MeV$, the maximum mass of the star can reach 2${M_ \odot }$ 
	
	\begin{table}[htbp]
		\centering
		\caption{We give the maximum masses and their
			corresponding radius for specific central energy density when $\alpha  =  - 3 \times {10^{ - 5}}$, $B = {B_ \bot }$, and ${a_4} \ne a_4^ \bot $. The compactness and redshift values are measured with a dimensionless quantity.} 
		\label{tab:1}
		\begin{tabular*}{\columnwidth}{@{\extracolsep{\fill}}cccccccc@{}}
			\hline  
			$ B={B_ \bot } $  & $ {a_4} $ & $ a_4^ \bot $ & $ {M_{\max }} $  &  $ R $ &  $ \rho _c $ & $ 2MG/R{c^2} $  & ${Z_{surf}}$ \\
			$ \mathrm{MeV}/\mathrm{fm}^3 $ &  &  & $  M_{\odot} $   & $ km $ &  $ \mathrm{MeV}/\mathrm{fm}^3 $ & &  \\
			\midrule[1.25pt]
			57 & 0.2 & 0.1 & 2.01 & 10.86 & 1330 & 0.545 & 0.482                                                \\
			&     & 0.3 & 1.96 & 10.80 & 1280 & 0.536 & 0.468                                                \\
			&     & 0.5 & 1.95 & 10.78 & 1270 & 0.533 & 0.464                                                \\
			&     & 0.7 & 1.94 & 10.78 & 1260 & 0.532 & 0.461                                                \\
			&     & 0.9 & 1.94 & 10.77 & 1260 & 0.531 & 0.460                                                \\
			& 0.5 & 0.1 & 2.05 & 10.98 & 1340 & 0.551 & 0.493                                                \\
			&     & 0.3 & 2.01 & 10.94 & 1280 & 0.542 & 0.478                                                \\
			&     & 0.5 & 2.00 & 10.92 & 1270 & 0.540 & 0.474                                                \\
			&     & 0.7 & 1.99 & 10.91 & 1260 & 0.538 & 0.471                                                \\
			&     & 0.9 & 1.99 & 10.90 & 1260 & 0.537 & 0.470                                                \\
			70 & 0.7 & 0.1 & 1.87 & 9.95  & 1730 & 0.553 & 0.496                                                \\
			&     & 0.3 & 1.83 & 9.92  & 1640 & 0.544 & 0.481                                                \\
			&     & 0.5 & 1.82 & 9.91  & 1620 & 0.541 & 0.477                                                \\
			&     & 0.7 & 1.81 & 9.90  & 1610 & 0.540 & 0.475                                                \\
			&     & 0.9 & 1.81 & 9.90  & 1600 & 0.539 & 0.473                                                \\
			92 & 0.9 & 0.1 & 1.64 & 8.65  & 3000 & 0.559 & 0.506\\
			&     & 0.3 & 1.61 & 8.69  & 2390 & 0.546 & 0.483                                                \\
			&     & 0.5 & 1.60 & 8.69  & 2320 & 0.542 & 0.478                                                \\
			&     & 0.7 & 1.59 & 8.68  & 2300 & 0.541 & 0.476                                                \\
			&     & 0.9 & 1.59 & 8.68  & 2280 & 0.540 & 0.474 \\
			\hline		
		\end{tabular*}
	\end{table}
	
\begin{figure}
		\centering
		\subfigure{
			\includegraphics[width=0.4\textwidth]{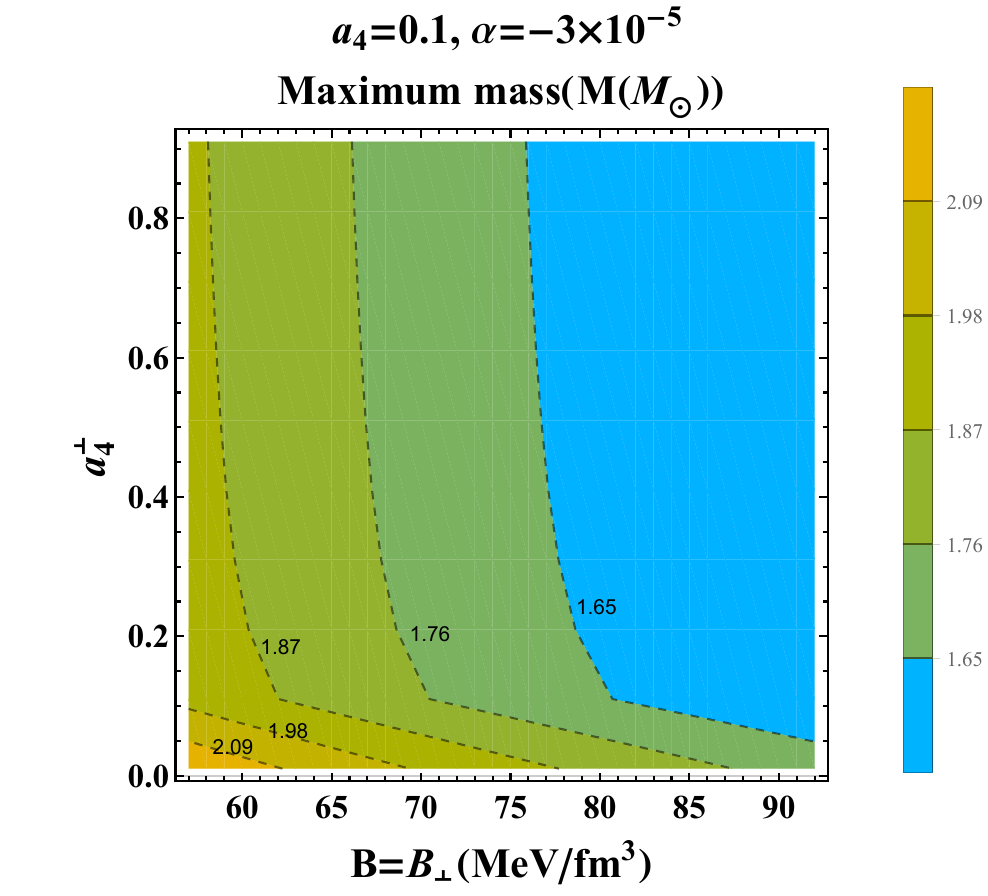}
			\includegraphics[width=0.4\textwidth]{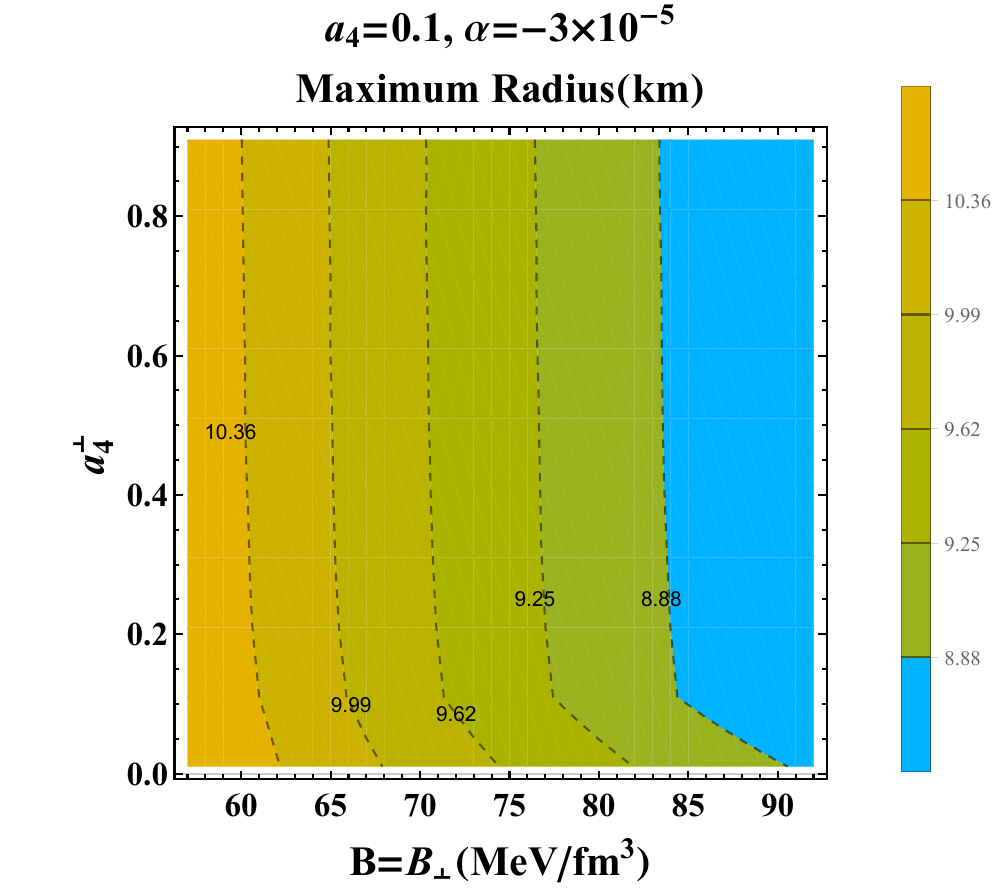}
		}
	\subfigure{
		\includegraphics[width=0.4\textwidth]{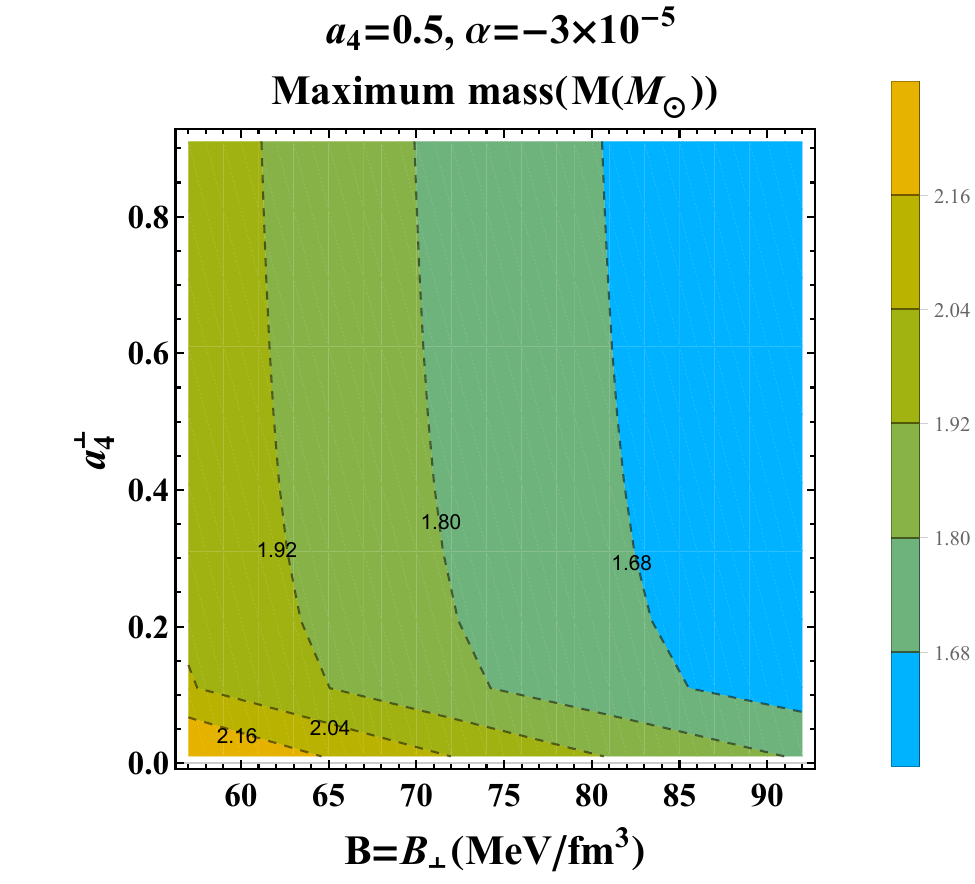}
		\includegraphics[width=0.4\textwidth]{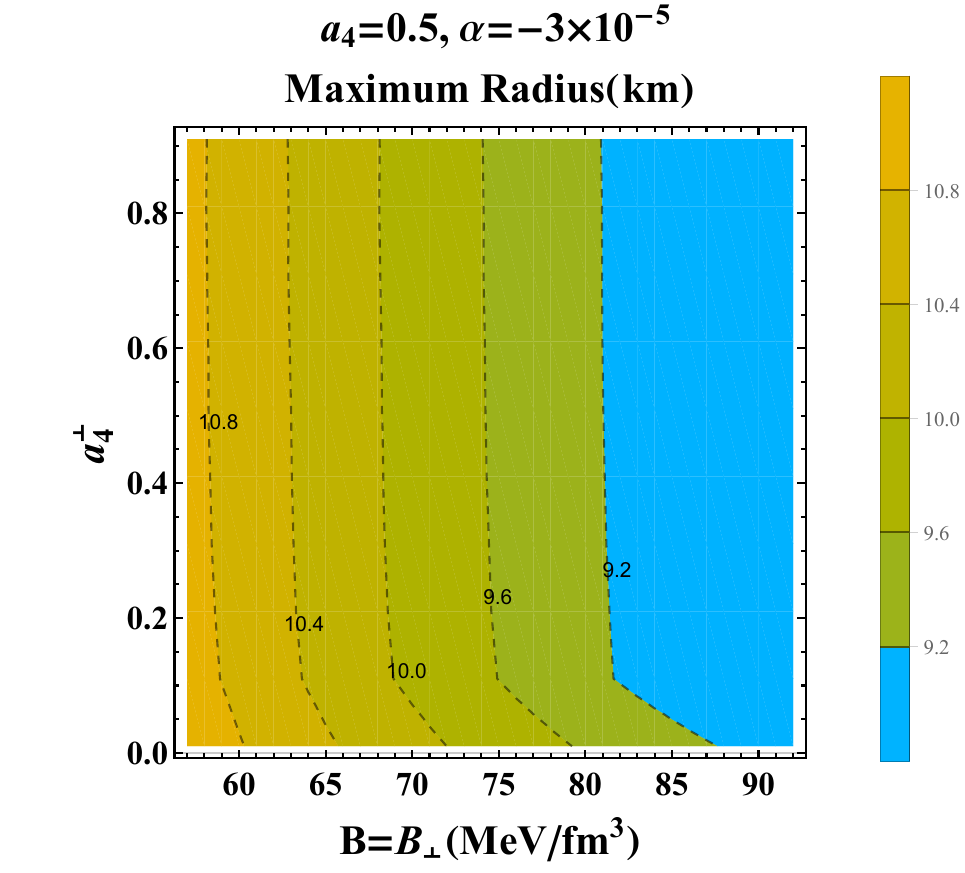}
	}
\subfigure{
	\includegraphics[width=0.4\textwidth]{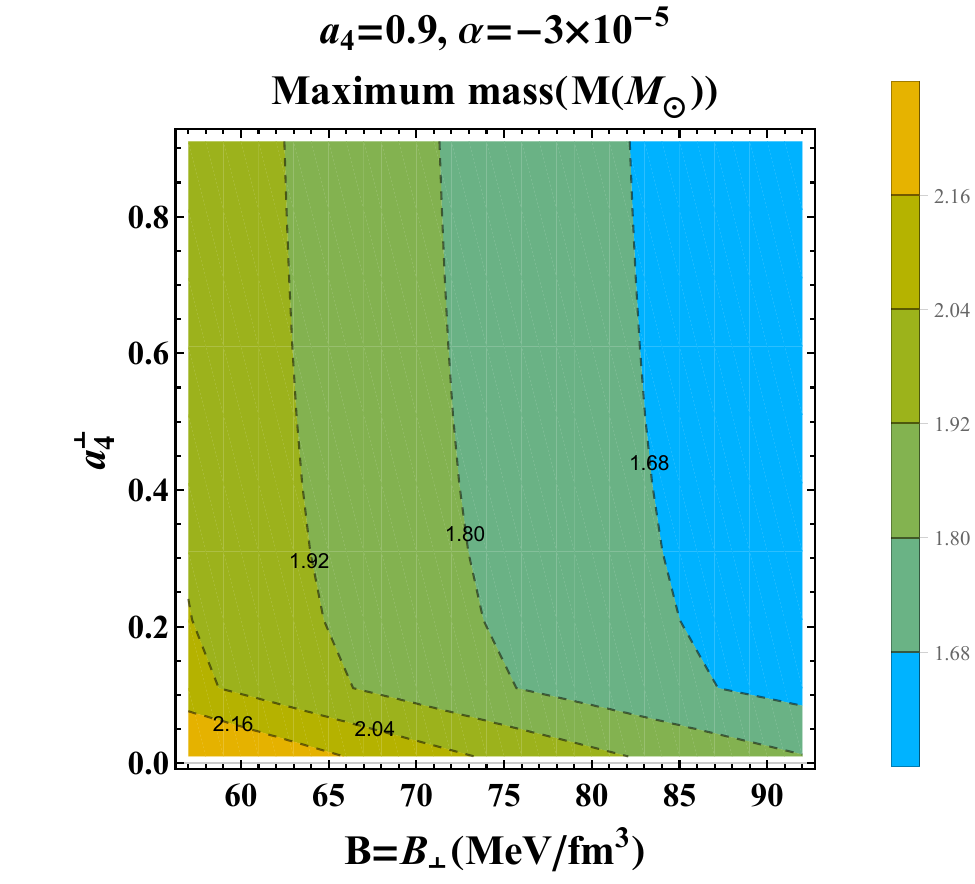}
	\includegraphics[width=0.4\textwidth]{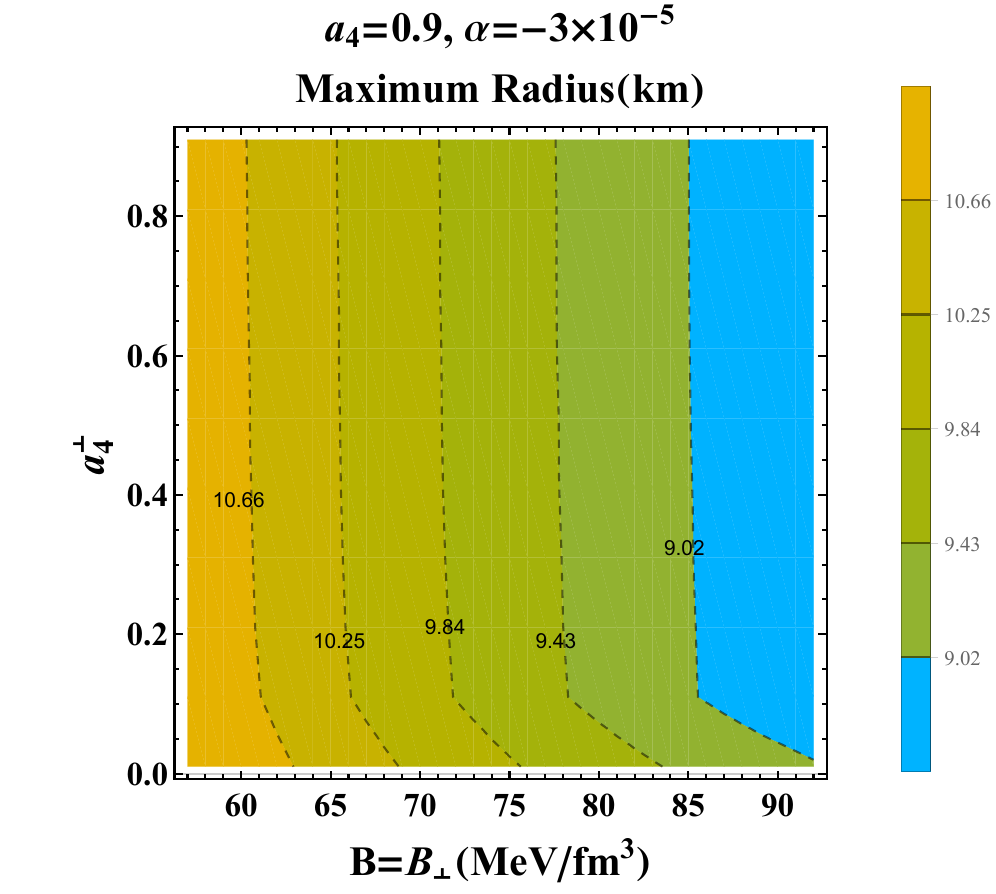}
}
		\caption{A contour plot depicting the maximum mass and its corresponding radius has been generated. At this point we set $57 < B = {B_ \bot } < 92MeV/f{m^3}$ and $0 < a_4^ \bot  < 1$. ${a_4}$ to be 0.1 (first row), 0.5 (second row) and 0.9 (third row), respectively. The black dashed lines are quark stars with the same maximum mass and their corresponding radii.} \label{fig:2}
	\end{figure}

	In Table \ref{tab:2}, we have established a set of parameters, with $B = 57MeV/f{m^3}$, ${B_ \bot } = 92MeV/f{m^3}$, ${m_s} = 50MeV$, ${a_4} = 0.9$ and $a_4^ \bot  = 0.1$, and corresponding configurations were applied for each value of $\alpha$, specifically$\alpha  =  - 3 \times {10^{ - 5}}, - 1.5 \times {10^{ - 5}},0,4 \times {10^{ - 5}}$. We have predicted the radius of the pulsar from a number of important astrophysical observations. In Table \ref{tab:2}, we give the quality of observations for PSR J0348+0432 \cite{antoniadis2013massive}, PSR J1614-2230 \cite{demorest2010two}, Vela X-1 \cite{gangopadhyay2013strange, rawls2011refined}, PSR J0437-4715 \cite{verbiest2008precision}, 4U 1820-30 \cite{guver2010mass}, PSR B1913+16 \cite{weisberg2005binary}, PSR J0030+0451 \cite{miller2019psr, riley2019nicer}, LMC X-4 \cite{rawls2011refined}, and SMC X-1 \cite{aziz2019constraining}, as well as the radius of the prediction for different values of parameter $\alpha$.
	 \begin{table}[htbp]
		\centering
		\caption{Radius predictions for PSR J0348+0432, PSR J1614-2230, Vela X-1, PSR J0437-4715, 4U 1820-30, PSR B1913+16, PSR J0030+0451, LMC X-4 and SMC X-1. }\label{tab:2}
		\begin{tabular}{|c|c|c|c|c|c|}
	\hline
	\multirow{3}{*}{candidates} & \multirow{3}{*}{\begin{tabular}[c]{@{}l@{}}$\begin{array}{*{20}{c}}\\ {Mass}\\ {M({M_ \odot })}\\ \end{array}$\end{tabular}} & \multicolumn{4}{l|}{Predicted Radius R(km)}\\ 
    \cline{3-6} & & \multicolumn{4}{l|}{$\alpha (f{m^3}/MeV)$}  \\ \cline{3-6} 
	&                                                                                                                                & \multicolumn{1}{l|}{$- 3 \times {10^{ - 5}}$}      & \multicolumn{1}{l|}{$ - 1.5 \times {10^{ - 5}}$}   & \multicolumn{1}{l|}{0}                             & $4 \times {10^{ - 5}}$        \\ \hline
	PSR J0348+0432              & $2.01 \pm 0.04$                                                                                                                & \multicolumn{1}{l|}{$11.47_{ + 0.05}^{ - 0.11}$}   & \multicolumn{1}{l|}{$11.41_{ + 0.694}^{ + 0.063}$} & \multicolumn{1}{l|}{$11.34_{ + 0.078}^{ - 0.237}$} & $-$                           \\ \hline
	PSR J1614-2230              & $1.97 \pm 0.04$                                                                                                                & \multicolumn{1}{l|}{$11.52_{ + 0.022}^{ - 0.05}$}  & \multicolumn{1}{l|}{$11.47_{ + 0.029}^{ - 0.063}$} & \multicolumn{1}{l|}{$11.42_{ - 0.078}^{ + 0.036}$} & $11.26_{ + 0.064}^{ - 0.155}$ \\ \hline
	Vela X-1                    & $1.77 \pm 0.08$                                                                                                                & \multicolumn{1}{l|}{$11.49_{ - 0.072}^{ + 0.046}$} & \multicolumn{1}{l|}{$11.46_{ - 0.071}^{ + 0.041}$} & \multicolumn{1}{l|}{$11.43_{ - 0.067}^{ + 0.036}$} & $11.34_{ - 0.058}^{ + 0.023}$ \\ \hline
	PSR J0437-4715              & $1.76 \pm 0.20$                                                                                                                & \multicolumn{1}{l|}{$11.49_{ - 0.232}^{ + 0.042}$} & \multicolumn{1}{l|}{$11.45_{ - 0.222}^{ + 0.026}$} & \multicolumn{1}{l|}{$11.42_{ - 0.216}^{ + 0.007}$} & $11.33_{ - 0.200}^{ + 0.052}$ \\ \hline
	4U 1820-30                  & $1.58 \pm 0.06$                                                                                                                & \multicolumn{1}{l|}{$11.28_{ - 0.091}^{ + 0.078}$} & \multicolumn{1}{l|}{$11.26_{ - 0.086}^{ + 0.077}$} & \multicolumn{1}{l|}{$11.23_{ - 0.085}^{ + 0.076}$} & $11.16_{ - 0.083}^{ + 0.072}$ \\ \hline
	PSR B1913+16                & $1.4408 \pm 0.008$                                                                                                             & \multicolumn{1}{l|}{$11.06_{ - 0.014}^{ + 0.014}$} & \multicolumn{1}{l|}{$11.04_{ - 0.014}^{ + 0.014}$} & \multicolumn{1}{l|}{$11.02_{ - 0.014}^{ + 0.014}$} & $10.95_{ - 0.014}^{ + 0.014}$ \\ \hline
	PSR J0030+0451              & $1.34_{ - 0.16}^{ + 0.15}$                                                                                                     & \multicolumn{1}{l|}{$10.87_{ - 0.349}^{ + 0.278}$} & \multicolumn{1}{l|}{$10.86_{ - 0.346}^{ + 0.268}$} & \multicolumn{1}{l|}{$10.83_{ - 0.348}^{ + 0.262}$} & $10.77_{ - 0.335}^{ + 0.259}$ \\ \hline
	LMC X-4                     & $1.29 \pm 0.05$                                                                                                                & \multicolumn{1}{l|}{$10.78_{ - 0.119}^{ + 0.093}$} & \multicolumn{1}{l|}{$10.75_{ - 0.107}^{ + 0.104}$} & \multicolumn{1}{l|}{$10.73_{ - 0.107}^{ + 0.104}$} & $10.67_{ - 0.095}^{ + 0.104}$ \\ \hline
	SMC X-1                     & $1.04 \pm 0.09$                                                                                                                & \multicolumn{1}{l|}{$10.16_{ - 0.245}^{ + 0.239}$} & \multicolumn{1}{l|}{$10.14_{ - 0.269}^{ + 0.238}$} & \multicolumn{1}{l|}{$10.14_{ - 0.268}^{ + 0.220}$} & $10.09_{ - 0.269}^{ + 0.220}$ \\ \hline
		\end{tabular}
	\end{table}	
	We further conducted research to explore the variations of maximum mass and its corresponding radius with respect to changes in $B = {B_ \bot }$ and $a_4^ \bot $. The numerical results obtained are represented using contour plots of $57 < B = {B_ \bot } < 92MeV/f{m^3}$ and $0 < a_4^ \bot  < 1$, with ${a_4}$ taking values of 0.1 (first row), 0.5 (second row) and 0.9 (third row), as depicted in Figure \ref{fig:2}. The research findings indicate that as the bag constant $B = {B_ \bot }$ increases, both the maximum mass and its corresponding radius exhibit a decreasing trend. As parameter $a_4^ \bot $ increases, the maximum mass and its corresponding radius also exhibit a decreasing trend, eventually stabilizing. Conversely, an increase in parameter ${a_4}$ leads to an increase in both the maximum mass and its corresponding radius.
		
	\section{Physical properties of stars other than M-R profiles}
	\label{sec:EMSG-g5}	
	In this section, we analyze the physical properties of the anisotropic quark star model and test them by means of graphing. We also test the stability of the internal structure of dense stars under EMSG theory.		
	\subsection{The stability criterion and adiabatic indices}	
\begin{figure}
	 	\centering
	 	\subfigure{
	 		\includegraphics[width=0.5\textwidth]{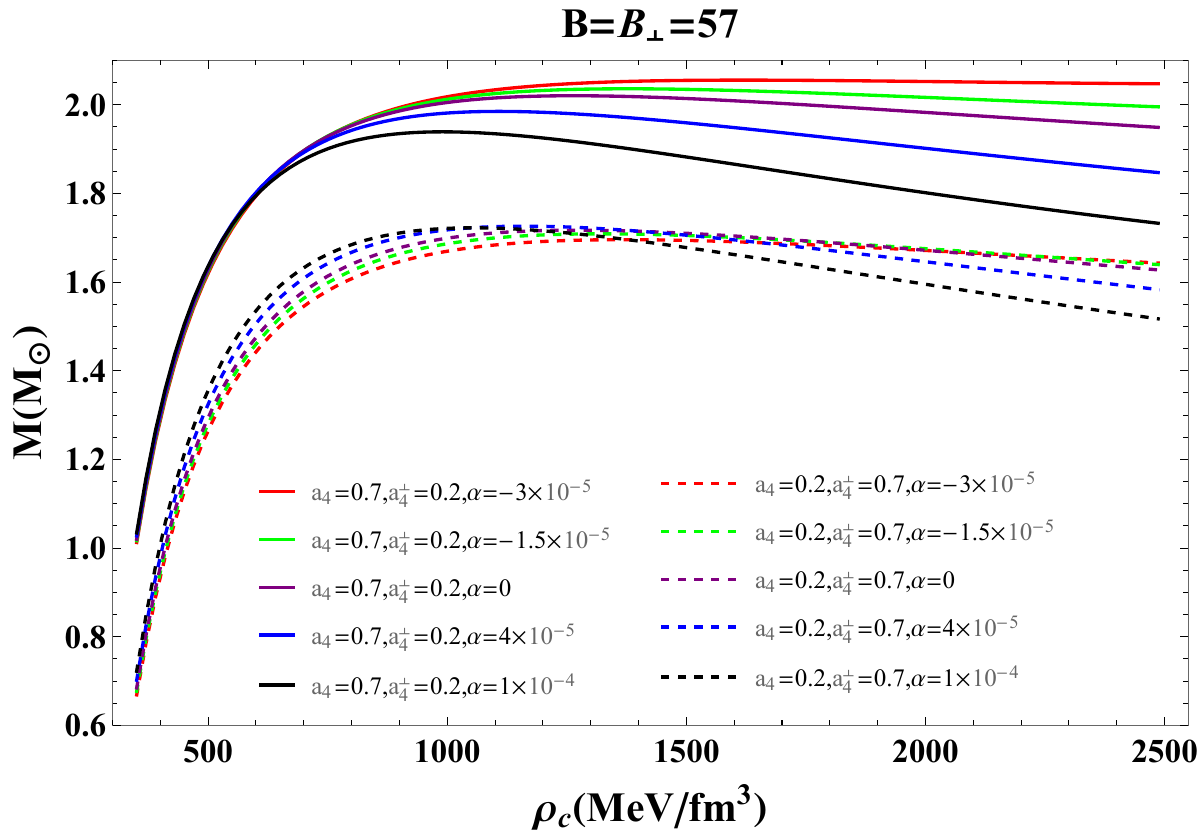}
	 		\includegraphics[width=0.5\textwidth]{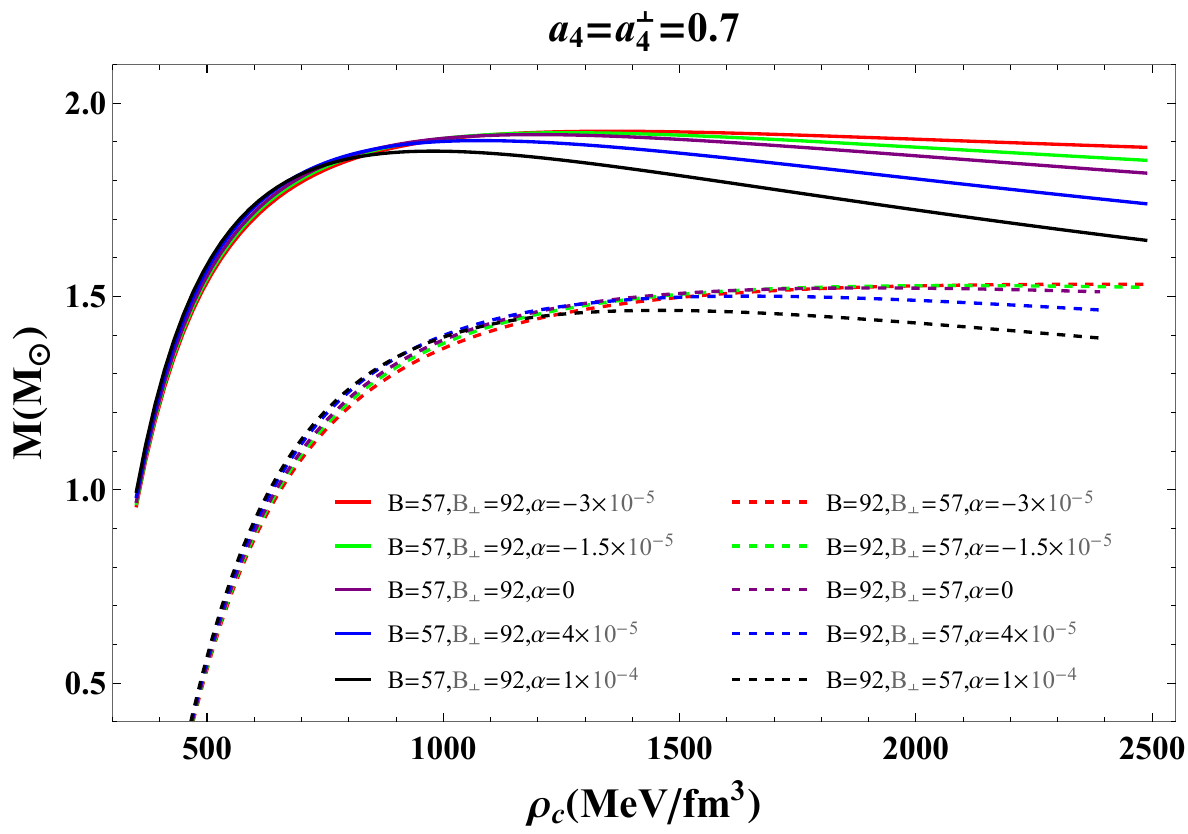}
	 	}
	 	\caption{A mass-central energy density relation diagram for anisotropic quark stars has been plotted. (i)$B = {B_ \bot } = 57MeV/f{m^3}$, ${a_4} \ne a_4^ \bot $ (left panel); (ii)${a_4} = a_4^ \bot  = 0.7$, $B \ne {B_ \bot }$ (right panel).} \label{fig:3}
	 \end{figure}

	Next, we can examine the stability of quark stars constructed under EMSG by analyzing the relationship between the total mass $M$ and central energy density ${\rho _c}$ at the $\frac{{dM}}{{d{\rho _c}}} = 0$ point. The Harrison-Zeldovich-Novikov criterion \cite{harrison1965gravitation, zeldovich1971relativistic} as	
	\begin{equation}
	\begin{split}
	&\frac{{dM}}{{d{\rho _c}}} < 0\begin{array}{*{20}{c}}
	{}&{ \to \begin{array}{*{20}{c}}
		{}&{}
		\end{array}}
	\end{array} unstable  \quad configuration\\
	&\frac{{dM}}{{d{\rho _c}}} > 0\begin{array}{*{20}{c}}
	{}&{ \to \begin{array}{*{20}{c}}
		{}&{}
		\end{array}}
	\end{array} stable \quad configuration
	\end{split} \label{Eq28}
	\end{equation}	
	This criterion applies to all quark stars structures. In Figure \ref{fig:3}, we have plotted the function depicting the relationship between the total mass $M$ of quark stars and the central energy density ${\rho _c}$. From the results in Figure \ref{fig:3}, it can be observed that only the initial portion of the curve, preceding the attainment of the maximum mass value, corresponds to a stable structure. Therefore, the points on the curve at the maximum mass value serve as a boundary point distinguishing between stable and unstable structures, achieving the differentiation between these two structural states.
	\subsection{Compactness and Redshift}	
\begin{figure}
		\centering
		\subfigure{
			\includegraphics[width=0.5\textwidth]{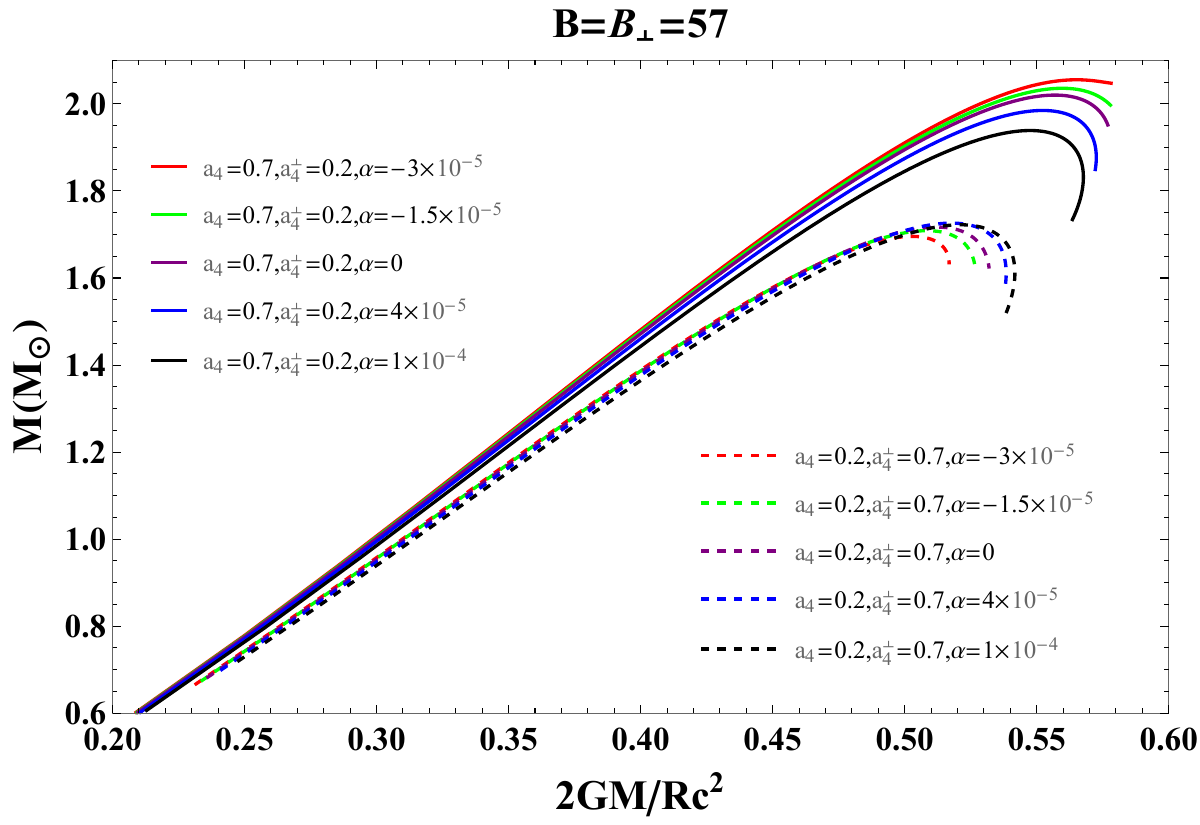}
			\includegraphics[width=0.5\textwidth]{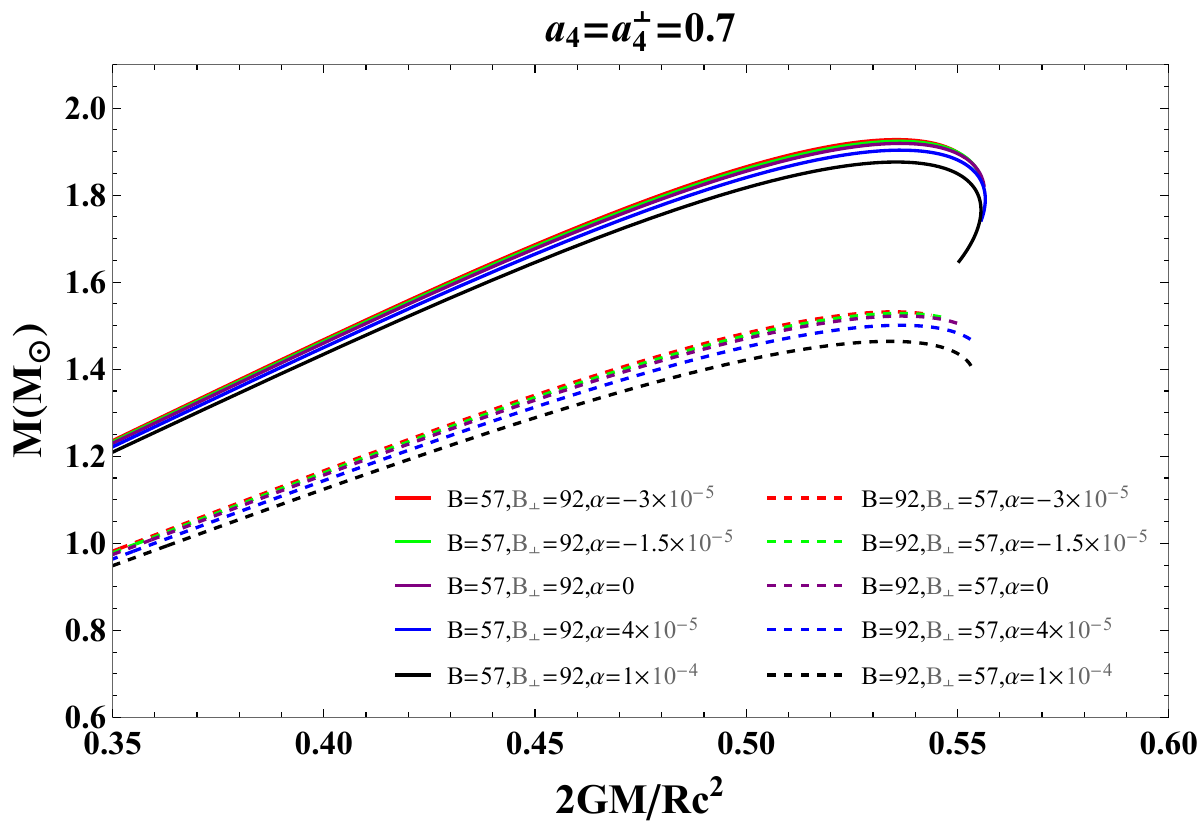}
		}
		\caption{Plots the image of the quark star Compactness $C = 2MG/R{c^2}$ as a function of the maximum mass M. (i)$B = {B_ \bot } = 57MeV/f{m^3}$, ${a_4} \ne a_4^ \bot $ (left panel); (ii)${a_4} = a_4^ \bot  = 0.7$, $B \ne {B_ \bot }$ (right panel).} \label{fig:4}
	\end{figure}

	To further explore the structure of quark stars, we also investigate the compactness $C = 2MG/R{c^2}$ of the EoS. In Figure \ref{fig:4}, we plot the image of the quark star compactness $C = 2MG/R{c^2}$ as a function of the maximum mass $M$. 
	In Figure \ref{fig:4}, we find that a maximum value of the densification factor $C$ occurs and satisfies the Buchdahl constraint i.e. $C \le 8/9$ \cite{buchdahl1959general} . Next the gravitational redshift can be defined by the factor of compactness as ${Z_{surf}} =  - 1 + {(1 - C)^{ - 1/2}}$ \cite{l2007neutron}. The gravitational redshift is related to the photons emitted from the surface of stars, making the study of gravitational redshift of significant importance to astronomers. In Table \ref{tab:1}, we have enumerated the gravitational redshift values corresponding to different numerical values of parameters $B = {B_ \bot }$, ${a_4}$, and $a_4^ \bot $, respectively.
	\section{Conclusions}
	\label{sec:EMSG-g6}	
	We investigated the influence of anisotropic pressure on compact stars within the framework of the EMSG theory. One crucial characteristic of this theory is that it no longer satisfies the conservation of energy-momentum tensor. We derived the modified TOV equation for compact stars and investigated the internal structure of quark stars under the assumption that they are composed of interacting quark matter, thereby considering the presence of anisotropy.
 	
	We performed a numerical solution of the modified TOV equation using the EoS for interacting quark matter, yielding the mass-radius relationship and other relevant physical properties of quark stars. In this process, we comprehensively considered the influence of different parameters on the quark star mass-radius relationship and compared it to actual observational data of pulsars. We utilized astronomical observations to predict the radii of pulsars based on their measured masses. We focused on the influence of parameters $a_4^ \bot $, ${a_4}$ and $B = {B_ \bot }$ on the maximum mass and its corresponding radius by generating contour plots of maximum mass and their respective radii. We observed that by reducing parameters $a_4^ \bot $ and $B = {B_ \bot }$ while simultaneously increasing parameter ${a_4}$, the maximum mass of quark stars can be increased to above 2.0${M_ \odot }$. For instance, when parameters $a_4^ \bot  = 0.1$, ${a_4} = 0.5$, and $B = {B_ \bot } = 57MeV/f{m^3}$, the mass of quark stars under the EMSG theory is approximately 2.05 ${M_ \odot }$, as detailed in Table \ref{tab:1}. We also find that by decreasing $a_4^ \bot $ when anisotropy is present will yield denser stars under EMSG theory than the isotropic case (see Table \ref{tab:1}). Finally, we confirm that the model is dynamically stable based on the variational approach by analyzing various physical properties of the dense stars. We conclude that within the framework of EMSG theory, considering anisotropy may result in denser stars compared to isotropic scenarios.\\	
 \section{Acknowledgments}
	\label{sec:EMSG-g7}
    Xian-MingLiu is supported by the NSFC Grant Nos.11711530645, 11365008 and the Program for Innovative Youth Research Team in University of Hubei Province of China (GrantNo.T201712). Jia-YongXiao is supported by Hubei Minzu University Graduate Innovation Research Key Project (MYK2020017). We thank  Defu Hou and Bingfeng Jiang for valuable discussions.

\bibliographystyle{unsrt}

\end{document}